\documentclass[prb,twocolumn,showpacs,preprintnumbers,amsmath,amssymb,notitlepage]{revtex4-1}
\usepackage{graphicx}
\usepackage{ulem}
\UseRawInputEncoding

\begin{document}

%\preprint{APS/cond-mat}
\title{ Instabilities in a running superfluid: boosted superfluid and stripe supersolid }
%\title{ Quantum Phase transitions observed in a moving frame and Galileo transformation near Quantum Phase transitions }
\author{  Jinwu Ye   }
\affiliation{ $^{1}$ Institute for Theoretical Sciences, Westlike University, Hangzhou, 310024, Zhejiang, China  \\
$^{2}$  Department of Physics and Astronomy, Mississippi State University, MS, 39762, USA     }
\date{\today }

\begin{abstract}
% Classical incommensurate phases and commensurate to in-commensurate transitions were studied in the physical systems of atom adsorption
% on periodic substrates. However, there are very little works on  their quantum counterparts.
% Classical incommensurate phases and commensurate to in-commensurate transitions were studied in the physical systems of adatom adsorption
% on periodic substrates. However, there are very little works on  their quantum counterparts.
%  The intrinsic interplay between the Galileo transformation and symmetry breaking in many body quantum systems are analyzed.
%  Implications for an anti-ferromagnet in a Zeeman field with $ z=2 $ and a  Ferromagnet in a staggered field  with $ z=1 $ are given.
  The possible instabilities in a running superfluid has been a long-time historical  problem since first studied by L. P. Landau.
  By constructing effective actions in terms of suitable order parameters, we revisit this outstanding  open problem at $ d=2 $ and $ d=3 $.
  We find that if the instability is driven by the SF Goldstone mode near $ k=0 $,
  then there is a quantum Lifshitz transition from the SF to a Boosted SF (BSF)  with
  the dynamic exponent $ (z_x=3/2,z_y=3) $  subject to logarithmic corrections from a marginally irrelevant cubic derivative term at $ d=2 $
  ( becomes irrelevant at $ d=3 $ ).
  This case may happen to exciton superfluids in 2d bilayer quantum Hall systems or electron-hole bilayer systems, especially
  in 2d weakly interacting Bose gas in cold atom systems.
  If the instability is driven by the roton mode near a finite momentum $ k=k_0 $, then there is a SF to a stripe supersolid  transition with
  the dynamic exponent $ z=1 $ which is in the same universality class as the $ z=1 $ boosted Mott-SF transition studied previously in a
  different context.
%  In the latter case, we argue that the resulting stripe solid may also likely have vacancies whose BEC may lead to a stripe supersolid.
  This case may apply to 3d Helium 4 and also  3d cold atom BECs with long-range interactions where the rotons in the SF phase plays an important role.
  Driving a SF sufficiently fast may become an effective way to create a SS which is a long time sought novel state of matter.
%  We also apply our approach to study possible new QPTs induced by directly driving a superfluid Helium4 with a given momentum
%  $ \vec{Q} $ and stress the roles of roton leading to a possible stripe supersolid.
\end{abstract}

%{\sl  Substantially revised version of cond-mat/0512480.  This project was initiated 17 years ago, could not finish until in July 15, 2022. It was inspired  by the %two recent joint works with Fadi Sun }

\maketitle

{\sl 1. Introduction: }
 It is well known that many systems become a superfluid at sufficiently low temperatures \cite{anderson}.
 He4 or He3 is the oldest strongly interacting bosonic or fermionic systems which become a SF below $ T_c \sim 2.17 K $ and
 $ T_c \sim 1 mK $ respectively. Since the discovery of the laser-cooling techniques,
 the weakly interacting superfluid systems were created in bosonic \cite{coldBEC1} and fermionic \cite{coldBEC2} charge neutral atoms
 at even lower temperature $ \sim nK $.
 There are also some experimental evidences to suggest exciton superfluid of electrons and holes may have been
 realized in the  2d bilayer quantum Hall systems in a strong magnetic field
 at the total filling factor $ \nu_T=1 $ \cite{BLQHwave} or  electron-hole bilayer at zero magnetic field
 with sufficiently long lifetime \cite{ehbl1}.
 In a quantum magnet subject to a Zeeman field, the $ U(1)_s $ spin rotation symmetry around the Zeeman field
 resembles the global $ U(1) $ symmetry in the interacting bosons \cite{z2}, the magnon condensations
 leading to some magnetic ordered phases can be mapped to the boson condensations leading to the SF.
% breaking the $ U(1)_s $ symmetry and leading to some magnetic ordered phases
% can be  mapped to boson condensations breaking the $ U(1) $ symmetry and leading to the SF phase.
% The the magnon condensations breaking the $ U(1)_s $ symmetry and leading to some magnetic ordered phases
% can be  mapped to boson condensations breaking the $ U(1) $ symmetry and leading to the SF phase.
 Recently, this mapping was also found in quantum magnets with spin-orbital couplings (SOC) in
 a longitudinal Zeeman field \cite{response}.

\begin{figure}[tbhp]
\includegraphics[width=.9 \linewidth]{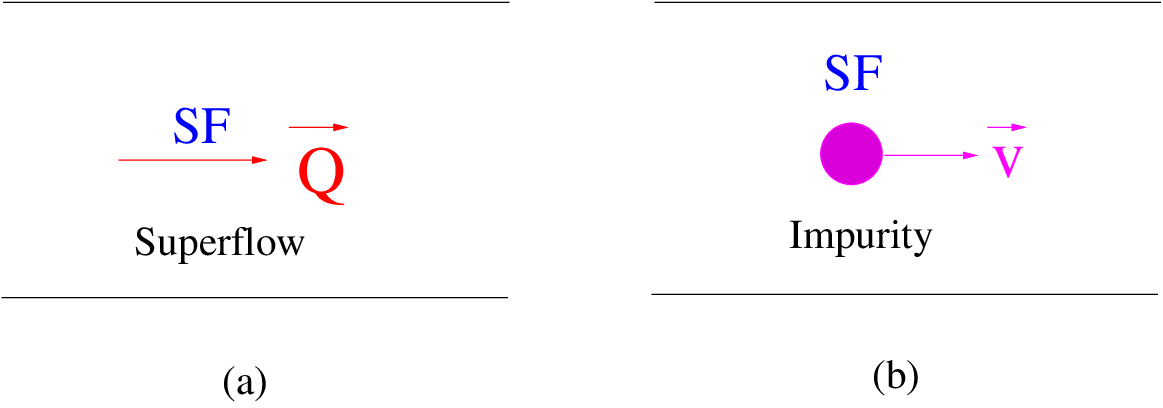}
\caption{ (a) Driving a superfluid relative to a straight wall.
     It is also equivalent to moving the straight wall relative to the SF.
 (b) A moving impurity in a superfluid which is different than  (a). }
\label{drivingfigs}
\end{figure}

 The physics of driving an object inside a  superfluid (SF) or driving the SF itself
 has a long history \cite{Landau,ring1,ring2,Fetter}.
 It may be necessary to distinguish the two different, but related cases:

 (1) Controlling the superfluid ( Fig.\ref{drivingfigs}a ):
 The SF is flowing with a finite velocity $ v $ with respect to a wide straight wall.
 It was also discussed in \cite{Landau,Fetter} and more recently in \cite{flowSF}.
 The flow of a SF with $ v > v^{SF}_c $
 may not destroy SF, but the order parameter may develop small additional components around a roton minimum, therefore reduce the
 superfluid density. However, when increasing $ v $ further, the fate of SF is still not known yet.
 (2) Controlling the moving object ( Fig.\ref{drivingfigs}b ):
 An impurity moving in a superfluid. It was discussed in \cite{Landau,Fetter} and more recently in \cite{impurity}.
 If an object moves in a superfluid at $ T=0 $ with a velocity below the critical velocity $ v < v^{O}_c $, there is no viscosity.
 However, when $ v > v^{O}_c $, a viscosity arises due to the emission of elementary excitations such as vortex rings \cite{ring1,ring2}.
 The two classes are completely different, so need separate discussions:
 the first class is an equilibrium steady one.
 The second  one is an non-equilibrium driven system which was experimentally investigated in cold atom BEC,
 by pulling an optical lattice\cite{pulling}.
 In this manuscript, we focus on the first class from effective action approach in either phase or
 dual density representation whichever is suitable. We will also establish some intrinsic connections between this class and
 a seemingly unrelated problem: the fate of liquid Helium under the increasing pressure investigated previously\cite{sscoll,ssrev,ss3}.
 While we relegate the discussions on the second class to the appendix B.

\begin{figure}[tbhp]
\centering
\includegraphics[width=0.4 \linewidth]{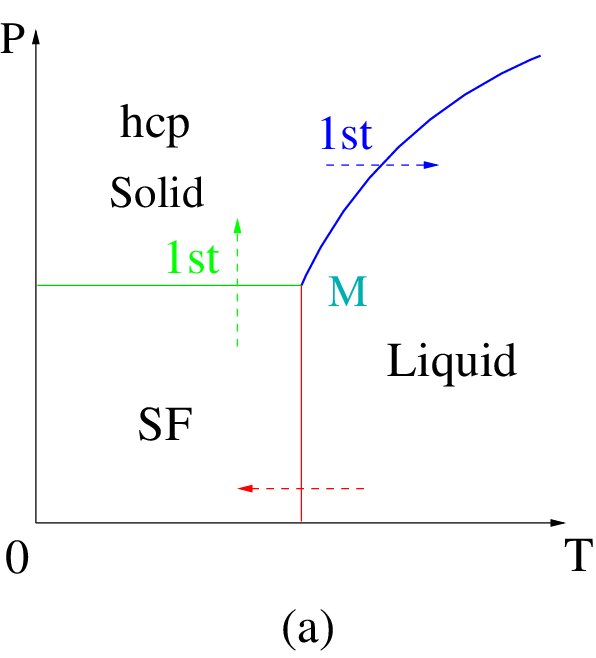}
\hspace{0.5cm}
\includegraphics[width=0.4 \linewidth]{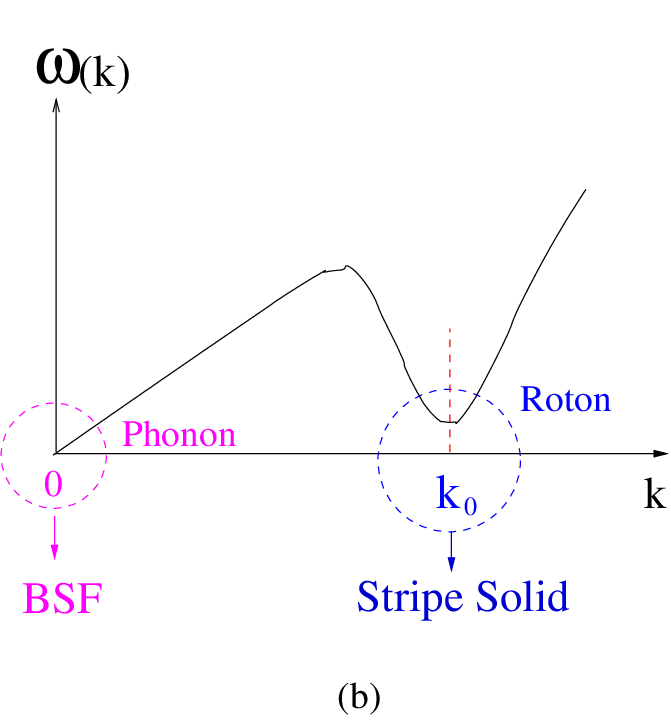}
\caption{
(a) The pressure $ P $ and temperature $ T $ He4 phase diagram with $ T_c \sim 2.13 K $ and $ P_c \sim120 bar $.
The liquid to SF transition is  in a classical 3d XY class for any pressure $ 0< P < P_c $ which is also exactly marginal.
The SF to solid transition is a first order quantum Lifshitz transition triggered by the lowering of roton surface
tuned by the pressure \cite{ss3}.   The liquid to solid transition is a first order classical Lifshitz transition
driven by the peak in the density-density correlation tuned also by the pressure.
A tantalizing possibility of a supersolid (SS) phase denoted by the dashed line
turns out to be just a phase separation due to the 1st order SF to solid transition.
(b) The elementary excitation in the SF phase. The roton surface is spherically symmetric
with the  roton gap $ \Delta/k_B \sim 10 K $ at the minimum  $ k_0 \sim 2 \AA^{-1} $.  }
\label{compareHe4}
\end{figure}

The global Pressure-Temperature (T-P ) phase diagram\cite{Fetter} of the Helium-4 is  shown in Fig.\ref{compareHe4}a.
The elementary excitations in the Helium-4 consist of the SF phonon part near $ k \sim 0 $ and
the roton part near $ k = k_0 $ ( Fig.\ref{compareHe4}b ).
%The roton surface is spherically symmetric.
In this well known T-P diagram, there maybe a room near the SF-Solid boundary  to host a possible tantalizing  supersolid
phase which has both the off diagonal SF order and the diagonal solid order
\cite{ssold}. In 2005, by using a torsional oscillator measurement, Chan's group \cite{chan} observed
a marked $1 \sim 2\% $ non-classical rotational inertial (NCRI) of the solid
4He at $∼ \sim 0.2K $ when $ P_c=120 bar  < P < 170 bar $,
both when embedded in Vycor glass and in bulk Helium 4. The authors suggested that the NCRI
may suggest the supersolid state of 4He. These experimental results inspired extensive
theoretical \cite{sscoll,ssrev} and experimental interests to examine  the very intriguing supersolid phase of 4He.
However, a later refined experiment \cite{laterHeexp} excludes the putative SS in Fig.\ref{compareHe4}a.
Despite its absence in the He4 system, the SS phase is an interesting phase on its own.
It was shown to exist in a lattice system \cite{gan,ss1L,ss2L,ss3L,honey,dual1,dual2}.
Of course, the SS in a lattice system is quite different than that in a continuum system \cite{ssrev}.
So a SS phase of matter in a continuum  remains a science fiction.

%    Of course, in the deep Mott limit $ U/t \rightarrow \infty $,
%    every optical lattice traps tightly integer number of bosons, then all the atoms in the Mott state move together with the lattice.
%    The center of mass ( or any given lattice point in the optical lattice ), just like a classical moving object, is Galileo invariant,
%    then the deep Mott state with its gap $ \Delta \rightarrow \infty $ responds trvially to the boost.
%   In the opposite deep SF limit $ U/t \rightarrow 0 $, the SF may response to the boost below the critical velocity similarly to (2) above.
%    This will be achieved in the present work.
%    On the other hand, if one performs the SF-Mott transitions in the moving frame where the optical lattice is at rest
%    and observed in the lab frame, we expect it remains the same.
%    We start from the microscopic model of boson-Hubbard model of interacting bosons at integer fillings
%in a square lattice which shows Mott to Superfluid (SF) transitions.
%In the continuum limit, the low energy effective actions have either the emergent Lorentz or the emergent Galileo invariance
%with the dynamic exponent $ z=1 $ and $ z=2 $ respectively.

In this manuscript, we will show that the SS may also exist in Helium 4 under a sufficiently large drive,
namely the second class of problem in the last paragraph. We develop an systematic and unified  effective action approach in both the phase and its dual magnitude representation to study all the possible instabilities and quantum phase transitions (QPT) by driving a SF beyond some critical velocities.
%Then we apply our formalism to study the effects of directly boosting the SF which leads to new classes of QPTs.
If the instability is due to the SF Goldstone mode near $ k=0 $, then there is a quantum Lifshitz transition
from the SF to a boosted SF (BSF)  with the novel dynamic exponent $ (z_x=3/2,z_y=3) $ and
a marginally irrelevant cubic derivative term at $ d=2 $.
We work out the excitation spectrum in both phases, perform the renormalization group analysis and
evaluate the scaling functions at a finite T near the QPT.
This scenario may also apply to the 2d exciton SF in BLQH \cite{BLQHwave} and ELBL \cite{ehbl1,ss2}
where the magneto-roton exists only at a very high energy, especially in cold atom BEC systems
which has very low critical velocities \cite{coldBEC1,coldBEC2}.
If the instability is due to the roton mode near $ k=k_0 $, then there is a SF to a stripe supersolid (SSS)  transition with
the dynamic exponent $ z=1 $ which is in the same universality class
of the boosted Mott-SF transition studied in \cite{response,moving} with an emergent C- symmetry.
We also analyze the symmetry breaking and the excitation spectrum in the SSS phase.
Then as the boost increases further, there is a QPT from the SSS to the stripe solid with $ z=2 $.
The resulting stripe solid may have vacancies whose BEC  leads to the SSS from the stripe solid side.
We show that the driving a superfluid is a new and effective mechanism to generate a supersolid which is a long time elusive goal
in low temperature physics and can be realized not only in He4, but also in various cold atom systems
with long-range dipole-dipole interactions, spin-orbit couplings or dressed by Rydberg atoms.
%This latter case may apply to Helium 4 where the rotons in the SF phase plays an important role.

% Implications for an anti-ferromagnet in a Zeeman field with $ z=2 $ and a  Ferromagnet in a staggered field  with $ z=1 $ are given.
% The intrinsic interplay between the symmetry breaking and the Galileo transformation are analyzed.

% which transform differently under the GT than the bare space-time in the microscopic Hamiltonian.
% As expected, our results in the SF side is consistent with those achieved by microscopic perturbation calculations
% in (2) ( also reviewed in Appendix D ) when the boost is below the critical velocity.
% But our methods can be applied to go beyond the critical velocity by finding the new phases and novel QPTs leading to these new phases.

{\sl 2. Co-moving frame and the lab frame in a running SF }

 In the co-moving frame with the SF, the SF is static. In terms of the SF order parameter $ \psi =\sqrt{\rho_0+\delta\rho}e^{i\phi}$,
 one just takes the effective action  inside the SF phase \cite{ss3}:
\begin{align}
	\mathcal{L}_{M:SF}[\delta\rho, \phi]& = i\delta\rho\partial_\tau\phi
	+\rho_0[v_x^2(\partial_x\phi)^2+v_y^2(\partial_y\phi)^2]  + u (\delta\rho)^2   \nonumber  \\
	 & + w \rho_0 ( \partial_y \phi )^3 + \cdots
\label{comovingSF}
\end{align}
where, in addition to phase-magnitude conjugation $ i\delta\rho\partial_\tau\phi $,
the last cubic derivative term also breaks the Charge conjugation ( C- ) symmetry
$ \phi \to -\phi, \delta \rho \to -\delta \rho $ in the $ ( \delta \rho, \phi) $ representation $ [\delta \rho, \phi]=i \hbar $.

 Now one gets to the lab frame just by performing a Galileo transformation (GT) \cite{response,moving}
 ( see also the appendix A for an independent microscopic calculation )
 $ \partial_\tau \rightarrow \partial_\tau -ic \partial_y $ ( we drop the $ \prime $  ):
\begin{eqnarray}
	\mathcal{L}_{L:SF}[\delta\rho, \phi]& = & i\delta\rho\partial_\tau\phi
	+\rho_0[v_x^2(\partial_x\phi)^2+v_y^2(\partial_y\phi)^2]  + u (\delta\rho)^2
	  \nonumber  \\
    & + &  w \rho_0 ( \partial_y \phi )^3  + c \delta\rho\partial_y \phi
\label{labSF}
\end{eqnarray}
  where the boost velocity was pinned to be $ \vec{c}= \vec{Q}/m $.

  The order parameter in the lab frame becomes
\begin{equation}
 \psi_{SF} =\sqrt{\rho_0+\delta\rho} e^{i ( \vec{Q} \cdot \vec{x} + \phi ) }
\label{Qmode}
\end{equation}
 which carries a SF flow in the lab frame.

 In the following, we will study the possible instability when the flow is beyond a critical one
 from both phase representation and its dual magnitude representation in the lab frame.
 The C- symmetry case was addressed in \cite{response} in a completely different context:
 a quantum magnet with SOC in a longitudinal Zeeman filed.
 The C- symmetry only exists in a lattice system at integer fillings \cite{boson0,response,moving}.
 However, there is no C- symmetry in all the continuous SF systems mentioned in the introduction.
 It is the absence of the C- symmetry which leads to the new QPTs in all the following sections \cite{Csymmetry}.

% As stressed below Eq.\ref{comovingSF}, the absence of C- symmetry plays a crucial role in
% all

{\sl 3. The quantum Lifshitz transition from the SF to the BSF  driven by the instability in the Goldstone mode }

 Taking $ \vec{c}= \vec{Q}/m $ as an independent tuning parameter, we will
 study the putative SF to the BSF transition tuned by this boost.
 see also the appendix A for an independent microscopic calculation.

  After integrating out the magnitude fluctuations in Eq.\ref{labSF}, the quantum phase fluctuations are described by:
\begin{align}
	\mathcal{L}_{L:SF}[ \phi] & =	\frac{1}{2 u }(\partial_\tau \phi-ic \partial_x \phi )^2
	+\rho_0[v_x^2(\partial_x\phi)^2+v_y^2(\partial_y\phi)^2]   \nonumber  \\
    &  + w \rho_0 ( \partial_y \phi )^3
\label{labSFphi}
\end{align}
  which has the Translational symmetry $ \vec{x}  \to \vec{x} + \vec{a} $ and
  the $ U(1)_c $ symmetry $ \phi \to \phi + \phi_0 $.

 Now we study how the SF evolves as one increases $ \vec{Q} $.
 The mean-field state can be written as $\phi=\phi_0+k_0 y$.
 Substituting it to the effective action Eq.\ref{labSFphi} leads to:
\begin{align}
    \mathcal{S}_{0}
	\propto( 2 u \rho_0 v_y^2-c^2)k_0^2+  2 w u \rho_0 k_0^3
\label{pmk0z2}
\end{align}

 At a low boost $c^2 < 2 U \rho_0 v_y^2 $, $k_0=0$ is in the SF phase which breaks the $ U(1)_c $ symmetry, but still keeps
 the translational symmetry. Its spectrum is given by:
\begin{align}
	\omega=\sqrt{2 u \rho_0(v_x^2k_x^2+v_y^2k_y^2)}-c k_y
\label{z2SFVc}
\end{align}

\begin{figure}[tbhp]
\centering
\includegraphics[width=0.4 \linewidth]{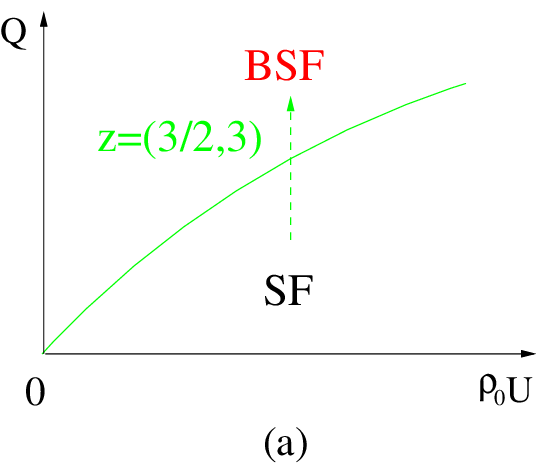}
\hspace{0.5cm}
\includegraphics[width=0.4 \linewidth]{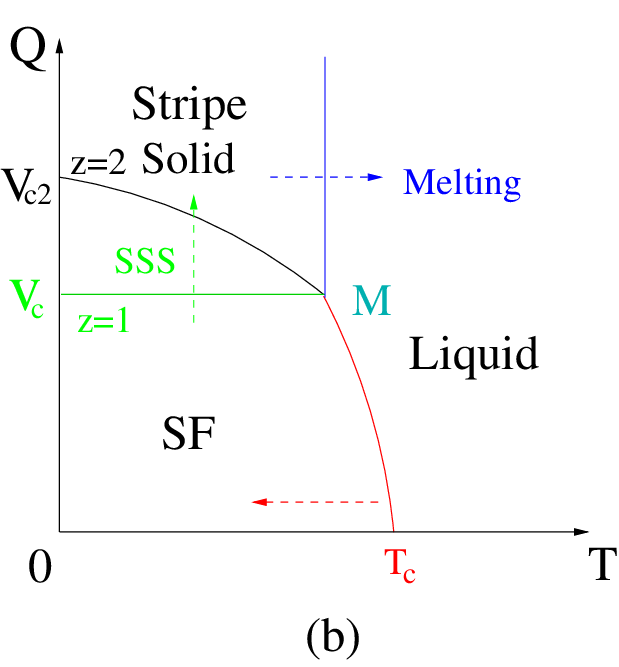}
\caption{ The instabilities of driving a SF.
(a)In the absence of a roton, the instability happens near the origin ( the phonon mode ), it leads to a BSF phase at $ T=0 $.
If there exists of a roton such as in He4, the instability at $ k=0 $ will always be pre-emptied by that near the roton at $ k=k_0 $.
(b) The instability near the roton minimum leads to a stripe supersolid (SSS) phase at $ T=0 $.
$ T_c \sim 2.13 K $ at $ Q=0 $ remains the same as in Fig.\ref{compareHe4}a, but decreases
as Eq.\ref{Tc} as $ Q $ increases. The critical velocity is estimated in Eq.\ref{cvaluenew} as $ v_c \sim 10 m/s $.
The QPT from the SF to the SSS is in the same universality class as that from the boosted Mott to the SF with $ z=1 $ \cite{response,moving}.
There is also a QPT from the SSS to the stripe solid with $ z=2 $ at $ v=v_{c2} $.
The finite temperature melting transition from the stripe solid to the normal liquid at $ T=T_M $ is also in the 3d XY universality class.
The  BEC of the vacancies in the stripe solid leads to the SSS intervening between the SF and the stripe solid.
There is an emergent C- symmetry at $ v= v_{c,d} $ dictating $ z=1 $, but no C- symmetry away from the QCP.
The order parameters in all the 4 phases are:
Normal liquid $ \langle \psi_0 \rangle= 0, \langle \psi_G  \rangle = 0 $,
SF $ \langle \psi_0 \rangle \neq 0, \langle \psi_G  \rangle = 0 $,
Stripe solid $ \langle \psi_0 \rangle = 0, \langle \psi_G  \rangle \neq 0 $,
SSS $ \langle \psi_0 \rangle \neq 0, \langle \psi_G  \rangle \neq 0 $. }
\label{drivingSF}
\end{figure}

  At the critical boost between the SF and the boosted SF
\begin{align}
	c^2=2U\rho_0 v_y^2=v^2_{c,p}
\label{z2right}
\end{align}
 which gives the phase boundary  in Fig.\ref{drivingSF}a.
 The physical meaning is clear: when the moving velocity of the SF matches that of the SF Goldstone mode.
 there is an instability to a new phase called boosted SF phase in the following.

At a high boost $   c^2 > 2 U \rho_0  v_y^2 $
\begin{equation}
   k_0 =\frac{c^2-2 U \rho_0 v_y^2 } { 3 w U \rho_0 }
\label{rhighboostz2}
\end{equation}
 where $ w = c $, so its sign is completely determined by the driving.
 So the BSF phase has an additional modulation $ k_0 $ along the $ y-$ axis on top of Eq.\ref{Qmode}:
\begin{equation}
 \psi_{BSF} =\sqrt{\rho_0+\delta\rho} e^{i [ ( \vec{Q} + \vec{k}_0 ) \cdot \vec{x} + \phi ] }
\label{QmodeBSF}
\end{equation}
 which still keeps the diagonal symmetry $ \vec{x} \to \vec{x} + \vec{a}, \phi \to \phi -\vec{k}_0 \cdot \vec{a} $.
 So the symmetry breaking is
\begin{equation}
   U(1)_T \times U(1)_c  \to [ U(1)_T \times U(1)_c ]_D
\label{u1u1}
\end{equation}
 which still leads to one Goldstone mode.

 Inside the BSF phase, the quantum phase fluctuations can be written as $ \phi \to \phi_0+k_0y+\phi $.
 Expanding the action upto the second order in the phase fluctuations leads to
\begin{eqnarray}
  \mathcal{L}_{BSF} & = & (\partial_\tau\phi-ic\partial_y\phi)^2
	+ 2 U \rho_0 v_x^2(\partial_x\phi)^2           \nonumber   \\
  & + & ( 2 c^2- 2 U \rho_0 v_y^2    )(\partial_y\phi)^2    \nonumber   \\
  & + & 2 w U \rho_0 (\partial_y\phi)^3 + b (\partial_y \phi)^4 + \cdots
\label{s2icdriving}
\end{eqnarray}
   which leads to the gapless Goldstone mode inside the BSF phase:
\begin{equation}
    \omega_\mathbf{k}=\sqrt{ 2 U \rho_0 v_x^2 k_x^2+( 2 c^2- 2 U \rho_0 v_y^2 )k_y^2 }-ck_y
\label{bsfdisdriving}
\end{equation}
where one can see  $ 2 c^2- 2 U \rho_0 v_y^2= c^2 + ( c^2- 2 U \rho_0 v_y^2 ) > c^2 $ when $c^2> 2 U \rho_0 v_y^2$,
thus the $\omega_\mathbf{k}$ is stable in BSF phase.

%{\sl 1. The exotic QCP scaling with the dynamic exponents $ (z_x=3/2, z_y=3 )  $ }

    It is instructive to expand the first kinetic term in Eq.\ref{labSFphi} as:
\begin{eqnarray}
    2 U \mathcal{L}
	 & = & Z(\partial_\tau\phi)^2
	-2i c \partial_\tau\phi\partial_y\phi
	+ 2 U \rho_0 v_x^2(\partial_x\phi)^2 + \gamma (\partial_y \phi)^2   \nonumber  \\
	 & +  & a(\partial_y^2\phi)^2
	+ 2w U \rho_0 (\partial_y\phi)^3  + b (\partial_y \phi)^4
\label{abz2}
\end{eqnarray}
  where $ Z $ is introduced to keep track of the
  renormalization of  $ (\partial_\tau\phi)^2 $, $ \gamma=2 U \rho_0 v^2_y- c^2= v^2_{c,p}- c^2 $ is the tuning parameter.
  By using the universal relation for the 2d  superfluid density  $ \Delta \rho_s /k_B T_c= 2/\pi $, one can deduce
  the finite temperature critical temperature \cite{lattice}:
\begin{equation}
    \frac{T_c}{T_{c0} }= | 1 - (\frac{c}{v_{c,p}})^2 | < 1
\label{Tc}
\end{equation}
  initial part of which at a small $ Q $  is shown in Fig.\ref{drivingSF}.

  The scaling $ \omega \sim k^3_y, k_x \sim k^2_y $ leads to the exotic dynamic exponents $ (z_x=3/2, z_y=3 )  $.
  Then one can get the scaling dimension of $ [\gamma]=2 $ which is relevant, as expected, to tune the transition,
  but $ [Z]=[b]=-2 < 0 $, so are two leading irrelevant operators which
  determine the finite $ T $ behaviours and corrections to the leading scalings. However $ [w]=0 $ is marginal.
  The standard field theory one-loop RG finds:
\begin{equation}
   \frac{ d w }{ d l}= \epsilon w - A w^3
\label{w22}
\end{equation}
 where $ \epsilon = 2 -d $ and  $ A=1/v^2_x a > 0 $.
 Note that the sign of $ w $ changes under the $ C- $ transformation $ \phi \to - \phi $.
 So Eq.\ref{w22} reduces to
\begin{equation}
   \frac{ d |w| }{ d l}= \epsilon |w| - A |w|^3
\label{w22ab}
\end{equation}

  So it is marginally irrelevant at $ d=2 $, simply irrelevant at $ d=3 $.
  Setting $ Z=w=0 $ in Eq.\ref{abz2} leads to the Gaussian fixed point action  at the QCP
  where $ \gamma=0 $, subject to the Logarithmic correction due to the marginally irrelevant $ w $ term.
  Again it is the crossing metric  $ g_{\tau, y}=g_{y, \tau}=-i c $ in Eq.\ref{abz2} which dictates the
  quantum  dynamic scaling near the QCP. It is a direct reflection of the new emergent space-time near the $ z=(3/2,3) $ QPT.
  Note that here the QPT is a quantum Lifshitz one tuned by $ [\gamma]=2 $, so despite the cubic derivative term $ [w]=0 $,
  it could still be a 2nd -order transition in Fig.\ref{drivingSF}a,
  in contrast to the conventional QPT where a cubic term drives a first order one.

 Now we evaluate the conserved currents in both SF and BSF phase.
 The former symmetry breaking pattern is just $ U(1)_c \to 1 $.
 The latter is  $ U(1)_T \times U(1)_c  \to  [ U(1)_T \times U(1)_c ]_D $.
 Due to the different symmetry breaking pattern, it is a QPT from the SF to the BSF.
 Now we try to find an order parameter to distinguish the two phases.
 The $ U(1)_c $ symmetry in the normal phase transpire
 as $ \phi \to \phi + a $ for any shift $ a $ inside the $ U(1)_c $ symmetry broken SF phase, so the Noether current can be
 derived from Eq.\ref{labSFphi} as:
\begin{eqnarray}
  J_\tau & = & 2( \partial_\tau \phi -i c \partial_y \phi )    \nonumber  \\
  J_x & = & 4 u \rho_0 v^2_x \partial_x \phi                   \nonumber  \\
  \tilde{J}_y & = & J_y-ic J_\tau=2u \rho_0[ 2 v^2_y ( \partial_y \phi ) + 3w ( \partial_y \phi )^2 ]-ic J_\tau
\end{eqnarray}

In the SF phase, $ \phi= \phi_0 $, then $ (J_\tau, J_x, \tilde{J}_y)= (0,0,0) $ and
$ (J_\tau, J_x, J_y)= (0,0,0) $ also. In the BSF phase, $ \phi= \phi_0 + k_0 y $ where
$ k_0 $ is given by Eq.\ref{rhighboostz2}, then  $ (J_\tau, J_x, \tilde{J}_y)= (-i 2c k_0, 0, 0 ) $,
but $ (J_\tau, J_x, J_y)= (-i 2c k_0, 0, 2c^2 k_0 ) $.
So the conserved currents $ (J_\tau, J_x, J_y) $ can still be used to distinguish the BSF from the SF phase.

%Using the $ \phi $ representation of Eq.\ref{boosttoright},
%one can also reproduce the current listed below Eq.\ref{z1current0} for the $ z=1 $ case.

However, if there exists roton shown in Fig.\ref{compareHe4}c, then this SF to BSF transition will be preempted by the
the SF to a solid transition triggered by the roton touchdown as shown in the following sections.

{\sl 4. The SF to the SF density wave transition driven by the instability in the roton mode }

 In Helium 4, due to the long-range Wan der-Waals interaction, the density-density interaction $ V_d(q) $ develops a roton minimum
 which drives the transition from the SF to a solid \cite{ssrev}.
 Now the density-density interaction $ U $ becomes long-ranged in Helium 4, so we adopt the notation in \cite{ss3}
 as $ U=V_d(k)= a -b k^2 + \alpha  k^4 $ which can be written as $ V_d(k)= \Delta + \alpha ( k^2- k^2_0 )^2 $
 near the roton minimum ( Fig.\ref{compareHe4}c ).
 We also consider the isotropic case $ v^2_x= v^2_y $, then
 $ \rho_s= \rho_0 v^2_x= \rho_0 v^2_y $ is the superfluid density,
 $ \kappa^{-1} = \lim_{k \to 0 } V_d(k) = a $ is the compressibility and  $ v^2(k)=  \rho_s V_d(k) $.

 From Eq.\ref{Qmode}, one can see that the density stays the same in both the lab frame and the co-moving frame.
 The dynamic structure factor is:
\begin{equation}
  S^{>}_n ( \vec{k}, \omega )=S_n(\vec{k}) \delta( \omega - \epsilon_{+}(\vec{k}) ),~~~S_n(\vec{k})= \frac{ \pi \rho_s k }{ 2 v(k) }
\label{structure}
\end{equation}
 where $ \epsilon_{+}(\vec{k}) = v(k) k + c k_x $ is the quasi-particle excitation energy.

 It is easy to see that a generalized  Feymann relation still holds under the driving:
\begin{eqnarray}
 \epsilon_{+}(\vec{k})= \frac{\int^{\infty}_0 d \omega  \omega  S^{>}_n ( \vec{k}, \omega )    }
 {  \int^{\infty}_0 d \omega  S^{>}_n ( \vec{k}, \omega )  }
\label{feymann}
\end{eqnarray}
 A similar relation for the quasi-hole  excitation energy $ \epsilon_{-}(\vec{k}) = v(k) k - c k_x $
 can be derived by replacing  $ S^{>}_n ( \vec{k}, \omega ) $
 by $ S^{<}_n ( \vec{k}, \omega ) $.

  After integrating out the phase fluctuations in Eq.\ref{labSF}, the quantum magnitude fluctuations
  can be used to describe such a transition under a driving:
\begin{align}
	\mathcal{L}[ \delta \rho] & =
	\frac{1}{2} \delta \rho ( - k, - \omega )
[   \frac{ \omega^2 + i 2 c \omega k_x }{ \rho_s k^2  } + ( \Delta- \frac{ c^2 k^2_x }{ \rho_s k^2 } )
               \nonumber  \\
  & + \alpha ( k^2-k^2_0 )^2 ] \delta \rho( k, \omega )
 -w ( \delta \rho )^3  + u ( \delta \rho )^4  + \cdots
\label{densitycubic}
\end{align}
  where $ r $ is the roton gap near $ k=k_0 $.
  Again the cubic term in the density-density channel need to be included at the very beginning.

  The $ U $ term nails down the momentum $ k $ to be in the roton ring  $ k=k_0 $,
  the boost term pins it to be in the $ k_x $ axis $ k_x= \pm k_0  $. So the boost term just introduce an easy-axis to
  the isotropic roton mode.
  So the resulting solid has only two shortest reciprocal lattice vectors $ \vec{G}= \pm k_0 \hat{x} $:
\begin{eqnarray}
   n & = & n_0 + ( \psi_G e^{i k_0 x} + \psi^{*}_G  e^{-i k_0 x} )    \nonumber  \\
     & = &  n_0 + 2 | \psi_G | \cos ( k_0 x + \alpha )
\label{decompG}
\end{eqnarray}
 where $ \psi_G $ is the complex order parameter. Its phase $ \alpha $ is the gapless phonon mode due to
 the translational symmetry breaking.
 It is the stripe solid phase. In fact,
 as shown in \cite{LOFF}, even without such an easy axis term which explicitly breaks
 the rotational symmetry, a strip solid phase is most likely to be the ground state lattice structure due to the spontaneously
 lattice symmetry breaking. In the presence of such an easy axis term, the stripe solid is the ground state.

  Then writing $ k_x= \pm k_0 + q_x, k_y=q_y $ and expanding up to the quartic term, we obtain:
 \begin{align}
	\mathcal{L}[ \delta \rho] & =
	\frac{1}{2} \delta \rho ( - \vec{q}, - \omega )
[   \frac{ \omega^2 + i 2 c \omega q_x }{ \rho_s k^2_0  }
   + 4 \alpha k^2_0 q^2_x + \frac{ c^2  }{ \rho_s k^2_0 } q^2_y    \nonumber  \\
   &  + \tilde{r}  ] \delta \rho( \vec{q}, \omega )
   -w ( \delta \rho )^3  + u ( \delta \rho )^4  + \cdots
\label{densitycubicq}
\end{align}
  where $ \tilde{r} = \Delta- c^2/\rho_s $ is the boosted roton gap and $ k_x= \pm k_0 + q_x, k_y=q_y $ need to be summed
  around both regimes near $  \pm k_0 $.
  Using the decomposition Eq.\ref{decompG}, one obtain the effective action describing the SF to the
  stripe solid transition \cite{qz}:
 \begin{align}
	\mathcal{L}[ \psi_G ] & =
	\frac{1}{2} \psi^{*}_G ( \vec{q}, \omega )
[   \frac{ \omega^2 + i 2 c \omega q_x }{ \rho_s k^2_0  }
   + 4 \alpha k^2_0 q^2_x +  \frac{ c^2  }{ \rho_s k^2_0 } q^2_y      \nonumber  \\
   & + \tilde{r}  ] \psi_G ( \vec{q}, \omega )
     + u | \psi_G |^4  + \cdots
\label{densitycubicpsiG}
\end{align}
  where due to the stripe structure, the cubic term plays no role.
  In fact as shown in \cite{LOFF}, the cubic term play an important role only in a triangular lattice
  where the three shortest reciprocal lattice vectors form a closed triangle.

  Substituting Eq.\ref{decompG} into the original boson Eq.\ref{Qmode}, we get the corresponding order parameter
  for the superfluid density wave (SDW):
\begin{equation}
 \psi_{SDW} =\sqrt{\rho_0} e^{i ( \vec{Q} \cdot \vec{x} + \phi ) }[1 +  | \psi_G | \cos ( k_0 x + \alpha )/\rho_0 ]
\label{SDW}
\end{equation}
  which establishes the relation between the original boson  $ \psi_{SDW} $ and the order parameter $  \psi_G $ in the effective action Eq.\ref{densitycubicpsiG}. The translational symmetry $ x \to x+ a $ in Eq.\ref{decompG} translates into
  the $ U(1)_T $ symmetry of $ \psi_G \to \psi_G e^{ i k_0 a }  $ where its phase $ \theta= k_0 a $ is any continuous real number \cite{CIC}.
  The $ U(1)_T $ symmetry breaking leads to
  the Goldstone mode which is the phonon mode $ \alpha $ in Eq.\ref{decompG} due to the translational symmetry breaking
  to the SDW phase.   It breaks the $ U(1)_T \times U(1)_c \to 1 $ symmetry leading to the two Goldstone modes $ \phi, \alpha $
  which are the  superfluid and lattice Goldstone mode respectively.

  Eq.\ref{densitycubicpsiG} can be rewritten as:
\begin{align}
	\mathcal{L}[ \psi_G ] & =
	\frac{1}{2} \psi^{*}_G ( \vec{q}, \omega )
[   \frac{  ( \omega + i  c  q_x )^2 + ( c^2 + 4 \alpha \rho_s k^4_0 ) q^2_x }{ \rho_s k^2_0  }   \nonumber  \\
   & +  \frac{ c^2  }{ \rho_s k^2_0 } q^2_y
    + \tilde{r}  ] \psi_G ( \vec{q}, \omega )
     + u | \psi_G  |^4  + \cdots
\label{densitycubicpsiGcomb}
\end{align}
  where $ c^2 <  c^2 + 4 \alpha \rho_s k^4_0 $, so it
  is nothing but in the same universality class of QPT from the boosted Mott to SF
  along the Path-I in Fig.3a in \cite{moving}.
  The boost $ c $ is exactly marginal, so it is a boosted 4D XY model with the dynamic exponent $ z_x=z_y=1 $.
  Despite the original action Eq.\ref{labSF} has no C- symmetry, it has an emergent C- symmetry in the effective action
  Eq.\ref{densitycubicpsiG}.

  Setting $ \tilde{r} = \Delta- c^2/\rho_s =0 $ leads to a new critical velocity:
\begin{equation}
  c^2= \rho_s \Delta= v^2_{c,d}
\label{cvalue}
\end{equation}
  which gives the SF to the SDW phase boundary  in Fig.\ref{drivingSF}b.
  When $ \tilde{r} > 0, \langle \psi_G \rangle =0 $, it is in the SF phase.
  When $ \tilde{r} < 0, \langle \psi_G \rangle \neq 0 $, it is in the Stripe SF density  wave (SFDW)  phase.
  It is new because it only depends on the roton gap $ \Delta $, independent of its minimum $ k_0 $.

  The main differences between the BSF in Eq.\ref{QmodeBSF} and the SDW in Eq.\ref{SDW} is that
  in the former, it is a phase fluctuation $ \phi $ driven QPT, so there is only one wavevector component $ \vec{Q} +\vec{k}_0 $.
  The symmetry breaking pattern is
  $ U(1)_T \times U(1)_c \to [U(1)_T \times U(1)_c]_D $ which leads to only one gapless Goldstone mode $ \phi $,
  while in the latter, it is a magnitude fluctuation $ \delta \rho $ driven QPT, so there are three wavevector
  components $ \vec{Q} $ and $ \vec{Q} \pm \vec{k}_0 $ which leads to the magnitude modulation in Eq.\ref{SDW}.
  The symmetry breaking pattern is $ U(1)_T \times U(1)_c \to 1 $ which leads to
  two gapless Goldstone modes $ \phi $ and $ \alpha $.

%  It can be contrasted to the naive one $ c_N= \Delta/k_0 $.  The correct one scales as $ \sqrt{\Delta} $,
%  the naive one as $ \Delta $.  This crucial difference should be subject to experimental tests in
%  driving He4 superfluid in Fig.\ref{compareHe4}d.

%  The scaling function in Sec.V-A of \cite{moving} applies here also.
%  The crucial difference is that this is the QPT from the SF to the strip solid tuned by $ c $ in Eq.\ref{cvalue}.

%  The question left is what will happen to the SF Goldstone mode near $ k=0 $ in Fig.\ref{compareHe4}b ?
%  Its fate determines if it is a stripe solid or stripe supersolid.

%  Of course, the SF owns its own Goldstone mode, but it remains un-critical throughout the
%  SF to the stripe solid transition, so was integrated out in Eq.\ref{densitycubicpsiG}.

%  at 3 dimension, $ u $ is marginally irrelevant, but $ w $ is relevant.
%  So the putative 2nd- order transition from the SF to the stripe solid at $ \tilde{r} = 0 $ is pre-emptied by the first
%  order transition triggered by the cubic term $ w $ before $ \tilde{r} $ goes down to zero.

{\sl 5. Crossover from the SDW to the stripe supersolid (SSS) and SSS to the stripe solid transition }

  Eq.\ref{SDW} holds near the SF to the SDW transition. As one increases the boost further,
  the superfluid density $ \rho_s $ starts to decrease as the normal solid component $ | \psi_G | $
  develops in Eq.\ref{decompG}. Then the density order Eq.\ref{decompG} emerges as an independent order parameter.
  The SDW crossovers to the stripe supersolid (SSS).
  One can write down a GL theory \cite{ss3} in terms of the two independent order parameters $ \psi $ and $ \delta n $ and their mutual couplings.
  For example, the periodic potential $ \delta n=  2 | \psi_G | \cos ( k_0 x + \alpha ) $ in
  Eq.\ref{decompG} acts as a periodic potential $ \delta n | \psi(x) |^2 $ on the superfluid component $ \psi $, so
  one can write down the generic expression for $ \psi $:
\begin{equation}
   \psi_{SSS}  =  \psi_0 e^{i \vec{Q} \cdot \vec{x}} [ 1 + A | \psi_G | \cos ( k_0 x + \alpha ) ]
\label{SSS}
\end{equation}
  where $ \psi_0 = \sqrt{a} e^{i \phi} $ and $ A $ is a numerical factor of the order 1.
  Well inside the stripe supersolid,  the coupling between the two gapless modes $ \phi $ and $ \alpha $
  leads to two branches of supersolidons \cite{ssrev}.

  As the boost increase further, the normal solid component $ | \psi_G | $ increases, the superfluid component
  $ \psi_{SSS} $ decreases and eventually disappears.
  There is a QPT from the SSS where
  $ \langle \psi_0 \rangle \neq 0, \langle \psi_G  \rangle \neq 0 $
  ( or equivalently $ \langle \psi_{SSS} \rangle \neq 0, \langle \delta n  \rangle \neq 0 $ )
  to the stripe solid $ \langle \psi_0 \rangle = 0, \langle \psi_G  \rangle \neq 0 $
  ( or equivalently $ \langle \psi_{SSS} \rangle = 0, \langle \delta n  \rangle \neq 0 $ ) in Fig.\ref{drivingSF}b.
  Just in terms of symmetry breaking, there is really no difference between the SDW and the SSS,
  so Eq.\ref{SDW} where $ a \sim \rho_0 $ and Eq.\ref{SSS} $ a \ll \rho_0 $ have the same symmetry breaking structure.
  But the former works best near the SF to the SDW transition
  described by Eq.\ref{densitycubicpsiG}  with $ z=1 $ where the SF component is non-critical
  ( so does $ \rho_s $ in Eq.\ref{densitycubicpsiG}  ). While the latter works best near the SSS to the stripe solid transition
  where the SF component becomes critical which is described by the $ z=2 $ effective action \cite{transfer}:
\begin{align}
  \mathcal{L}[ \psi_0 ] & =
  \psi^{*}_0 ( \vec{q}, \omega ) [ i Z_1 \omega  + v^2_x q^2_x +  v^2_y q^2_y
    - \tilde{\mu}  ] \psi_0 ( \vec{q}, \omega )     \nonumber  \\
     & + u | \psi_0 |^4  + \cdots
\label{vacancy}
\end{align}
 where $ \tilde{\mu}=v_{c2}-v $.
 It breaks the C-symmetry explicitly. The phenomenological parameters $ Z_1, v^2_x, v^2_y $ depend on
 $ Q $ and $ k_0 $ in Fig.\ref{drivingSF}b. Unfortunately, one is not able to determine the numerical value of
 such $ v_{c2} > v_c $ from the present effective action approach.
 At any finite $ T < T_{c} $ in Fig.\ref{compareHe4}b, the transition from the SSS to the stripe solid becomes a classical 3d XY
 one driven by the SF Goldstone mode $ \phi $. As the temperature increase further to $ T_c $, there is a second transition from
 the stripe solid to the normal liquid described by Eq.\ref{densitycubicpsiGnormal}. It is also a classical 3d XY
 one, but driven by the phonon mode $ \alpha $.

  It is constructive to compare with the extended boson Hubbard model \cite{dual1,dual2}
  where a stripe supersolid was shown to always exist slightly away from $ 1/2 $ filling both by microscopic calculations
  \cite{gan,ss1L,ss2L,ss3L,honey} and effective actions in the original basis \cite{gan,ss1L,ss2L,ss3L} and the dual vortex basis \cite{dual1,dual2}.
  Due to the lack of C- symmetry,
  the vacancies usually have lower energies than that of interstitials, the stripe solid always host some vacancies.
  If their excitation energies $ E_v $ are positive, so can only be thermally excited.
  But when they become negative, they may undergo BEC.
%  The emergent C- symmetry in Eq.\ref{densitycubicpsiG} may still disappear due to these vacancies.
  So approaching from the stripe solid side, the $ \psi_{SSS} $ in Eq.\ref{SSS} can be interpreted as  the BEC of vacancies
  in the spontaneously formed stripe solid in Eq.\ref{decompG}.
  The vacancies behave similarly as the holes on the top of the CDW or Valence Bond (VB)
  state at $ 1/2 $ filling examined in \cite{dual1,dual2} described by the effective action Eq.\ref{vacancy}
  with $ z=2 $.
%  It is their BEC which leads to the CDW-SS or VB-SS.
%  We expect it to be in the same class as the CDW to CDW-SS or VB to VB-SS  transition \cite{dual1,dual2}
%  with the dynamic exponent $ z=2 $.

{\sl 6. Normal Liquid to stripe solid transitions. }

%  As the boost increases further in Fig.\ref{drivingSF}b. The density of the vacancies gets smaller and smaller, then disappears
%  at the SS to the stripe solid transition.
  In fact, by only keeping the $ \omega=0 $ component in Eq.\ref{densitycubicpsiG}, it may also be used to describe
  the normal liquid to stripe solid transition at a finite $ T $ in Fig.\ref{drivingSF}b.
 \begin{align}
	\mathcal{L}[ \psi_G ] & =
	\frac{1}{2} \psi^{*}_G ( \vec{q}, \omega )
[   u^2_x q^2_x + u^2_y ( q^2_y +  q^2_z )  + \tilde{r}  ] \psi_G ( \vec{q}, \omega )      \nonumber  \\
   & + u | \psi_G  |^4  + \cdots
\label{densitycubicpsiGnormal}
\end{align}
  where we used the 3d-space explicitly \cite{qz}.
%  The gap $ \Delta $ is the one in the density-density correlation function ( the static structure factor )
%  $ S(k) \sim \frac{1}{ \Delta + \alpha (k^2 -k^2_0)^2 }  $  in the normal liquid.
  When $ \tilde{r}= T - T_c > 0, \langle \psi_G \rangle =0 $, it is in the normal liquid phase.
  $ \langle n \rangle =n_0 $.  $ \tilde{r} < 0, \langle \psi_G \rangle \neq 0 $,
  it is in the Stripe solid  phase  still described by Eq.\ref{decompG}
  $ \langle n \rangle   =   n_0 + 2 | \psi_G | \cos ( k_0 x + \alpha ) $.
%  Eq.\ref{cvalue} still holds.  The boost $ Q $ still plays an important role.
  It is in the 3d XY universality class.
  From the low temperature stripe solid side, it may also be understood as a stripe lattice melting transition driven by
  the phonons $ \alpha $ from the low temperature stripe solid side which also
  leads to Eq.\ref{densitycubicpsiGnormal}. While, as stressed below Eq.\ref{vacancy}, the finite $ T < T_c $ transition
  resulting from Eq.\ref{vacancy} is a 3d class driven by the SF Goldstone mode $ \phi $.

{\sl 7. Experimental realizations. }

  We study two kinds of instabilities: the one induced by the SF Goldstone mode near $ k=0 $ in the absence of rotons
  and the one induced by the roton mode near $ k=k_0 $.
  Here we discuss their experimental realizations respectively.

  Eq.\ref{z2right} shows that when the moving velocity of the SF matches that of the SF Goldstone mode.
  there is an instability to the BSF.
  The sound velocity in He4 is about $ v_{SF} \sim 238 m/s $.
  In a conventional lab on the earth, taking a high way ( magnetic levitated ) train moving with
  a velocity $ 300 km/h \sim 83 m/s $ is still below this characteristic velocity.
  A civil air-craft flight can reach even higher $ 800 km/h \sim 240 m/s $ which just reaches the sound velocity in the Helium4.
  So the chance to see the instability near $ k=0 $ is unlikely on the earth.
  As mentioned in the introduction,  in addition to the well known SF in the He4,
  there are also excitonic SF in the electronic systems in semi-conductors.
  The Bilayer Quantum Hall systems (BLQH) \cite{BLQHwave} hosts the exciton SF
 in the charge neutral sector  with the Goldstone mode velocity $ v_{BL}\sim 1.4 \times 10^{4} m/s $,
 The electron-hole bilayer system (EHBL)\cite{ehbl1} holds the exciton SF  with $ v_{EH}\sim 5 \times 10^{3} m/s $.
 It is essentially impossible except going to a satellite orbiting around the earth.
 The escape velocity of a satellite orbiting around the earth is $ v_{esc} = 11.2 km/s $.
 In this regard, the weakly interacting cold atom BEC systems \cite{coldBEC1,coldBEC2} become better candidates with
 the SF Goldstone mode velocity $ v_{SF} \sim 1 cm/s $.

   Eq.\ref{cvalue} shows that when the moving velocity of the SF is beyond a critical velocity determined by the roton gap,
   there is an instability to the stripe SS.  Its value can be estimated as:
\begin{equation}
   v_{c,d}= \sqrt{\frac{ \Delta }{2 m } }   \sim  10 m/s
\label{cvaluenew}
\end{equation}
  where $ m $ is the He4 atom mass and $ \Delta/k_B \sim 10 K $ is the roton gap.

  As mentioned in the introduction Fig.\ref{drivingfigs}a and reviewed in the appendix B-1,
  there is another critical velocity for an impurity moving in a superfluid.
  Its value due to the rotons in the He4 can be estimated by a simple Landau argument to be
  $ v_{imp} \sim \frac{\Delta}{ \hbar k_0} \sim 60  m/s $
  where $ k_0 \sim 2 \AA^{-1} $ is the momentum of the roton minimum.
  Our new critical velocity Eq.\ref{cvaluenew} holds for a uniformly moving SF which is equivalent to
  a uniforming moving channel. Obviously a moving local heavy impurity is different from a moving channel holding the SF.
  Just from the general principle, we expect the critical velocity of the cases are different, but comparable
  $ v_{imp} \sim v_{c,d} $. Our concrete calculations show that this is indeed the case due to
  $ \Delta \sim \frac{ \hbar^2 k^2_0 }{ 2 m } \sim 20 K $.

  It was also known that the critical velocity for the SF moving in a narrow channel with the width $ d $
  can be empirically fitted to be $ v_{ch} \sim \frac{\hbar}{md } $.
  We expect this form breaks down in an extremely narrow channel $ d \to 0 $ and
  also an extremely wide  channel $ d \to \infty $. Our new critical velocity Eq.\ref{cvaluenew}
  holds for extremely wide  channel $ d \to \infty $

  The SSS can be detected by the same NCRI experiments \cite{chan,laterHeexp}.
  The NCRI is proportional to the superfluid density which is, of course,
  anisotropic $ \sim a^2 v^2_x $ if the rotational axis is along the boost direction $ \hat{x} $,
  $ \sim a^2 v^2_y $ if the rotational axis is normal to the boost direction $ \hat{x} $.
  It will monotonically increases when moving from the SSS side near the $ z=2 $ line to the SDW side with $ z=1 $.

  Therefore Fig.\ref{drivingSF}
  can be mapped out \cite{chan,he4G} by driving the SF beyond the critical velocity in Eq.\ref{cvaluenew}.
  Cold atom BEC systems with a long-range interaction such as
  a dipole-dipole interaction \cite{dipole1,dipole2,dipole3,dipole4} or with spin-orbital couplings \cite{soc1,soc2}
  or dressed by Rydberg atoms \cite{Ryd1,Ryd2} may support
  rotons with much smaller critical velocities.
  So Fig.\ref{drivingSF} may find wide applications in these cold atom systems with tunable roton gaps.

{\sl 8. Conclusions.   }

  Obviously, the SF and the stripe solid phase break two completely different symmetries: In the the former, it is the internal
  $ U(1)_c $ symmetry whose breaking leads to the gapless Goldstone  mode,
  In the latter, it is the translational symmetry whose breaking leads to the gapless lattice phonon modes.
  Just from general symmetry principle, there are two possibilities on the QPT from the SF to the solid:
  (1) a direct 1st order transition. This is the case driven by the pressure $ P $ in He4 in Fig.\ref{compareHe4}a.
   The roton minima is rotationally symmetric.
   The QPT near $ k=k_0 $ is first order one driven by the pressure resulting a hcp solid structure,
   then the SF just disappears suddenly across the QPT.
   Of course, the first order transition indicates that  there could also be a coexistence of SF and solid near the SF-Solid phase boundary
   resulting a phase separation. This is indeed the case confirmed by the refined PSU experiment \cite{laterHeexp}.
   In fact, as a retrospect, the SS phase, despite it tantalizing properties,
   should not be expected in such an environment in the very first place.
  (2) It splits into two second order ones with an intervening SS phase.
  As demonstrated in this manuscript, this is the case in the 3d He4  or cold atom BEC with tunable rotons driven by the boost in Fig.\ref{drivingSF}b.
  The QPT near $ k=k_0 $ is second order one with $ z=1 $ driven by the boost $ Q $ resulting a stripe solid component,
  then the SF undergoes an accompanying second order QPT  to a  superfluid density wave (SDW).
  The boost transfers the rotationally symmetric roton minima to just two opposite degenerate minima along the boost direction.
  The QPT is in the same universality class as the $ z=1 $ boosted Mott-SF transition studied previously in a different context \cite{moving}.
  Then the Stripe SDW evolves into the stripe supersolid phase where
  the solid component is given by $ \delta \rho $, the supefluid density wave component is given by
  $ \psi_{SDW} $. Finally the SSS phase gets into a stripe solid phase through the $ z=2 $ QPT
  by kicking out of its SDW component $ \psi_{SDW} $ which stands for vacancies near the QPT from the stripe solid side.
%  To get a supersolid which hosts the coexistence of both the superfluid density-wave component and a solid component,
%  both have the same reciprocal lattice vectors, this coexistence was ruled out in the pressure ( P ) driving He4 by a more
%  refined experiment \cite{laterHeexp}. This is because the P driving is roton dropping 1st order transition resulting a
%  hcp solid structure (Fig.\ref{compareHe4}). Here it maybe possible due to its second order transition nature and
%  the solid component is given by $ \delta \rho $, the supefluid ( density wave ) component is given by
%  its phase $ \phi $. The $  (\delta \rho, \phi ) $ is  a pair of quantum mechanically conjugate variables.
  A microscopic calculation such as a QMC simulation
  is needed to confirm the existence of these vacancies and their stabilities against the phase separations.
%  If so, it could be called a stripe supersolid.
  We expect that phase separations are usually associated with the first order transition such as in  Fig.1a in the case (1),
  but not the second order one such as in Fig.2b in the case (2).
  If so, we may have discovered a new mechanism to realize a stable supersolid phase.
  Then the boost becomes an effective way to generate a
  supersolid which does not happen when increasing the pressure

This work was originally initiated by  the author's earlier works \cite{ss3,ssrev} on putative SS in He4.
I thank Moses Chan for helpful discussions during the very early stage of this work.
I also thank Fadi Sun for the collaborations on the two related works \cite{response,moving} which inspired the author to get back to
finish this early work.

%WE also thank Mark Novotny for proof-reading the abstract and introduction.
%We thank Prof. Gang Tian and Prof. Congjun Wu  for the hospitality during their visit at the West Lake university;
%also thank Prof. Wei Ku for the hospitality during their visit at the Tsung Dao Lee Institute in Shanghai, China.

\appendix

\section{ A moving superfluid, Doppler shifts and Galileo transformation }

 In this appendix, we review the topic (2) mentioned in the introduction.
 we perform some microscopic Bogliubov calculations to quadratic order  on a moving superfluid
 to demonstrates the known phenomena of Doppler shifts due to the Galileo transformation.
 They are consistent with the mean field + Gaussian fluctuation  analysis on the effective action approach used in the main text.
 However, the main limitation of the  microscopic calculations used in this appendix is that it is not able to
 determine what is the quantum phase beyond the critical velocity, let alone the universality class of the quantum phase transitions
 driven by the boost. One must push this Bogliubov calculations to infinite order to address this questions.
 This can only be achieved by the  effective action and RG approach used  in the main text.

{\sl 1. Doppler shift in  a moving SF at weak coupling: }

   The Hamiltonian for weakly interacting bosons in a continuum system such as the ESF in the
   EHBL \cite{ss2,ehbl1,ehbl2,ehbl3} is
\begin{equation}
     H_{B}= \sum_{\vec{k}} ( \epsilon_{\vec{k}} - \mu )  b^{\dagger}_{ \vec{k} } b_{ \vec{k} }
          + \frac{1}{2 A }  \sum_{\vec{k},\vec{p},\vec{q} }  V (\vec{q} )
          b^{\dagger}_{\vec{k} - \vec{q} } b^{\dagger}_{\vec{p} + \vec{q} }b_{\vec{p} }b_{\vec{k} }
\label{bosoncontin}
\end{equation}
   where $ \epsilon_{\vec{k}}= \hbar^2 k^2/2m $ is the free boson dispersion,
   $ \mu $ is the chemical potential, $ A $ is the 2d area, $ V_d (\vec{q} ) $ is the boson-boson interaction.
   We assume it is weak, so the following Bogliubov method applies.

   Setting the SF moving with a momentum $ \vec{Q} $:
\begin{equation}
   \psi_0 = \sqrt{N_0} e^{i \vec{Q} \cdot \vec{x} }
\label{Qtake}
\end{equation}
   which means a finite superflow $ \vec{v}= \vec{Q}/m  $ where $ m $ is the mass of an atom.
   Now one can write the boson operator as:
\begin{equation}
   \psi_{\vec{Q} + \vec{k} } = \sqrt{N_0} \delta_{ \vec{k},0} + b_{\vec{Q} + \vec{k} }
\label{away}
\end{equation}
 where $   b_{\vec{Q} + \vec{k} } $ stands for the quantum fluctuations with the momentum  $ \vec{k} $
 measured relative to the BEC momentum $ \vec{Q} $.

% In fact,  Eq.\ref{boson}  with the on-site interaction $ U $  can formally also be mapped to
%   Eq.\ref{bosoncontin} with $ \epsilon_{\vec{k}} =-2t( \cos k_x + \cos k_y ) $ and $ V_d (\vec{q} ) = U $
%   where $ \vec{k} $ is confined in the first BZ $ -\pi/a < k_x, k_y < \pi/a $. So the following Bogliubov calculation may also apply
%   to the strong SF limit $ U/t \ll 1 $ in Eq.\ref{boson}, but will fail near the QPTs in Fig.\ref{phasesz2}.
%   For an infinite-range interaction $  V_d (\vec{q} ) = U \delta_{\vec{q},0} $.
%   Eq.\ref{Qtake} and Eq.\ref{away} can be contrasted to the corresponding terms in a lattice Eq.\ref{k0phase}
%   and Eq. \ref{awaylattice}.

   Substituting Eq.\ref{away} into Eq.\ref{bosoncontin} and expanding it to the quadratic order,
   one can determine the chemical potential $ \mu $ by  eliminating the linear term of $ b_{\vec{q}} $ in the Hamiltonian $ H_{SF} $ as
\begin{equation}
   \mu =  n_0 V_d(0) + \frac{1}{2} m v^2
\end{equation}
   where $ n_0= N_0/A $ is the condensate density. Then one obtain the mean field Hamiltonian  $ H_{SF} $  to the quadratic order:
\begin{eqnarray}
    H_{SF}&= &\sum_{\vec{k}} [ \epsilon_{\vec{k}} + n_0 V_d ( \vec{Q}- \vec{k} ) - \epsilon_{\vec{Q}} ]  b^{\dagger}_{ \vec{k} } b_{ \vec{k} }
                                         \nonumber  \\
          &+ & \frac{ n_0}{2 }  \sum_{\vec{k}} [ V_d (\vec{k} ) b^{\dagger}_{\vec{Q} + \vec{k} }b^{\dagger}_{\vec{Q} - \vec{k} } + h.c. ]
\end{eqnarray}
   which can be diagonalized by  the Bogoliubov transformation
\begin{equation}
   \beta_{\vec{k}} =  u_{\vec{k}} b_{\vec{Q} + \vec{k} } + v_{\vec{k}} b^{\dagger}_{\vec{Q} - \vec{k} }
\end{equation}

   We obtain  $ H_{SF} $  in  terms of the quasi-particle creation and annihilation operators $ \beta_{\vec{k}} $
   and $ \beta^{\dagger}_{\vec{k}} $:
\begin{equation}
   H_{SF} =  E(0) +  \sum_{\vec{k}} E_v (  \vec{k} )\beta^{\dagger}_{\vec{k}} \beta_{\vec{k}}
\end{equation}
   where $ E(0) $ is the ground state energy and
\begin{eqnarray}
     u^2_{\vec{k}}  =  \frac{ \epsilon_{\vec{k}} + n_0 V_d ( \vec{k} ) }{ 2 E(  \vec{k} ) }+ \frac{1}{2}  \nonumber  \\
     v^2_{\vec{k}}  =  \frac{ \epsilon_{\vec{k}} + n_0 V_d ( \vec{k} ) }{ 2 E(  \vec{k} ) } - \frac{1}{2}  \nonumber  \\
     E_v  ( \vec{k} )   =  E(  \vec{k} ) + \vec{k} \cdot \vec{v}~~~
\label{uvE}
\end{eqnarray}
  where  $ E(  \vec{k} )= \sqrt{  \epsilon^2_{\vec{k}} + 2 n_0 V_d ( \vec{k} ) \epsilon_{\vec{k}} }  $.

  One can see that in the moving SF, $ u_{\vec{k}} $ and $ v_{\vec{k}} $ are the same as in the lab frame.
  However, the energy spectrum  $ E_v (  \vec{k} ) $ contains a Doppler shift term $ \vec{k} \cdot \vec{v} $
  \cite{dopplerhigh,thermalhigh}.  In the low energy limit $ \vec{k} \rightarrow 0 $ limit,
  $ E_v (  \vec{k} ) \rightarrow u | \vec{k} | + \vec{k} \cdot \vec{v}  $ where $ u= \hbar \sqrt{ \frac{n_0 V_d(0) }{m} } $ which is identical
  to Eq.\ref{z2right}.
  If one picks up $  \vec{k}  || - \vec{v} $, one can identify the critical velocity $ v_c= u $.
  Unfortunately, as stressed in the first paragraph, one is not able to tell what will happen beyond the critical velocity from
  this Bogliubov approach.
  This outstanding problem can only be addressed by the  effective action and RG approach demonstrated in the main text.

  One can also obtain normal and anomalous Green function:
\begin{eqnarray}
    G_n( \vec{Q}; \vec{k}, \omega ) & = & i \frac{ \omega -  \vec{k} \cdot \vec{v} + \epsilon_{\vec{k}} + n_0 V_d ( \vec{k} ) }
      { ( \omega -  \vec{k} \cdot \vec{v} )^2 -E^2 (  \vec{k} ) }    \nonumber  \\
     G_a( \vec{Q}; \vec{k}, \omega ) & = & i \frac{ n_0 V_d ( \vec{k} ) }
      { ( \omega -  \vec{k} \cdot \vec{v} )^2 -E^2 (  \vec{k} ) }
\label{Glab}
\end{eqnarray}
   where one can identify the excitation spectrum $ \omega= \pm  E(  \vec{k} ) + \vec{k} \cdot \vec{v} $
   which is nothing but the last equation in Eq.\ref{uvE}.

{\sl 2. Galileo transformation on the SF }

   The Galileo transformation is:
\begin{eqnarray}
      \vec{k}^{\prime} & = &  \vec{k}   \nonumber  \\
      E^{\prime}       & = & E- \vec{k} \cdot \vec{v} +  \frac{1}{2} m v^2  \nonumber  \\
      \mu^{\prime}     & = &  \mu + \frac{1}{2} m v^2
\label{kEmu}
\end{eqnarray}
  where $ \vec{k}^{\prime} , E^{\prime}, \mu^{\prime} $ are the momentum, energy and chemical potential in the moving frame,
  where $  \vec{k}, E , \mu $ are those in the lab frame. Note that the momentum does not change,
  because as listed in Eq.\ref{away}, it was measured from the BEC momentum  $ \vec{Q}= m \vec{v} $ from very beginning.

   In the moving frame, the Green functions take the same form as those in the lab frame:
\begin{eqnarray}
    G_n(  \vec{k}^{\prime}, \omega^{\prime} ) & = & i \frac{ \omega^{\prime} + \epsilon_{\vec{k}^{\prime}} + n_0 V_d ( \vec{k}^{\prime} ) }
      {  \omega^{\prime 2 } -E^2 (  \vec{k}^{\prime} ) }    \nonumber  \\
    G_a(  \vec{k}^{\prime}, \omega^{\prime} ) & = & i \frac{ n_0 V_d ( \vec{k}^{\prime} ) }
      {  \omega^{\prime 2 } -E^2 (  \vec{k}^{\prime} ) }
\label{Gmoving}
\end{eqnarray}
    By substituting
\begin{equation}
     \omega^{\prime} = E^{\prime}-\mu^{\prime}= E-\mu - \vec{k} \cdot \vec{v} = \omega - \vec{k} \cdot \vec{v}
\label{nonDoppler}
\end{equation}
    which is just the non-relativistic $ c_l \rightarrow \infty $ limit of a relativistic Doppler shift,
    one can see  Eq.\ref{Gmoving} recovers Eq.\ref{Glab}.

    It would be interesting to look at how the photons are emitted from the moving ESF in the EHBL systems \cite{ss2,ehbl1,ehbl2,ehbl3}, especially across the QPT from the SF to the BSF in Fig.\ref{drivingSF}a at both $ T=0 $ and a finite $ T $.
    As warned above, the Bogliubov method here may not be applied near any QPTs. One may extend the effective action developed in
    Sec.2 in the main text
    to couple to a photon bath to achieve this goal.
    This approach can also be applied to study the chiral edge state of a FQH phase \cite{chen}.

\section{ Driving a classical object through a superfluid  }

  In the last two subsections, we drive the quantum fluids, no classical objects embedded inside it.
  In this section, we study the case of driving a classical object through a continuous fluid sketched in
  Fig.\ref{drivingfigs}a.  It is a different class than that discussed in the main text, because here has both a classical object such as an impurity and the quantum fluid, while there is only quantum fluid in the latter.
  By analyzing how the GT act differently in the two cases, we also  stress the crucial differences
  than  the problems addressed in the main text.
  When above a critical velocity, these classical objects will all cause viscosities and dissipations.
  The three kinds of classical objects: a point impurity,  an underlying optical lattice or a straight wall
  correspond to different space symmetry: a spherical, translation by a lattice constant or translational symmetry along the wall,
  so should lead to different critical velocities.  Unfortunately, we are not able to provide any concrete solutions to such a different class of problems but  we outline possible approach to solve such a different class of problems.
  As long as there are relative motions between the quantum fluids and the classical objects,
  it does not matter if it is the classical object moving or the quantum fluid moving, both need to the same physics.
  So in the following, we assume it is the classical object which is moving.

   We start from a general Hamiltonian
\begin{eqnarray}
  {\cal H} = \int d^2 x \psi^{\dagger}( \vec{x} )
   [  -\frac{\hbar^2}{ 2 m} \nabla^2 + V_1( \vec{x} ) - \mu ] \psi( \vec{x} ) ~~~~~~~~~~~~~
                  \nonumber  \\
   +  \int d^2 x_1 d^2 x_2  \psi^{\dagger}( \vec{x}_1 )\psi( \vec{x}_1 )
   V_2(x_1-x_2 ) \psi^{\dagger}( \vec{x}_2 )\psi( \vec{x}_2 )  ~~~~~
\label{Seq0Nlattice2}
\end{eqnarray}
  where the single-body lattice potential $ V_1( \vec{x} ) $ and the two-body interaction $ V_2(x_1-x_2 ) $ are automatically
  incorporated into the kinetic term and the interaction term respectively.
  By adding the chemical potential $ \mu $, we also change the canonical ensemble with a fixed number of particles
  $ N $ in the first quantization to
  the grand canonical ensemble in the second quantization.
  In the following, to simplify the notation, we use $ V_{int}=\int d^2 x_1 d^2 x_2  \psi^{\dagger}( \vec{x}_1 )\psi( \vec{x}_1 )
   V_2(x_1-x_2 ) \psi^{\dagger}( \vec{x}_2 )\psi( \vec{x}_2 ) $.

{\sl 1. A moving impurity }

   In a continuous system with a translational invariance, $ V_1(x)=0 $ in Eq.\ref{Seq0Nlattice2}.
   We look at the case of driving the impurity with a given velocity $ v $ in the Hamiltonian Eq.\ref{Seq0Nlattice2}:
\begin{eqnarray}
  {\cal H}^i_L & = &  \int d^2 x \psi^{\dagger}( \vec{x} )
   [  -\frac{\hbar^2}{ 2 m} \nabla^2  - \mu- g_i \delta( \vec{x}- \vec{R}+ \vec{v} t ) ] \psi( \vec{x} )
   \nonumber  \\
   & + &  V_{int}
\label{driving}
\end{eqnarray}
  where $ \vec{R} $ is the initial position, $ \vec{v} $ is the velocity of the impurity, $ g_i $ is the scattering potential strength.
  As expected, it is a time-dependent driving system in the lab frame.

  In the frame moving together with the impurity \cite{RW}, one can setting $ \vec{x}^{\prime}= \vec{x}+ \vec{v} t $.
  Then one obtain the Hamiltonian in this co-moving frame ( still drop the $ \prime $ for the notational simplicity ):
\begin{eqnarray}
  {\cal H}^i_{M} & = & \int d^2 x \psi^{\dagger}( \vec{x} )
   [  -\frac{\hbar^2}{ 2 m} \nabla^2  - \mu- g_i \delta( \vec{x}- \vec{R} ) -i v \partial_x ] \psi( \vec{x} )
   \nonumber  \\
        & + &  V_{int}
\label{drivingprime}
\end{eqnarray}
  which, as expected, becomes time-independent in the moving frame.
%  The most dramatic difference between Eq.\ref{hhhh}   with $ V_1(x ) $  and the current case is that
%  in the former one boost both the quantum and classical degree of freedoms at the same speed,
%  so the relative distance between the boson and the impurity
%  $ \vec{x}^{\prime}- \vec{R}^{\prime}= \vec{x}- \vec{R} $ is invariant under the GT, so can be set to be zero in the tight-binding limit
%  in any inertial frame, so it is a time-independent problem in any inertial frame.
%  but here $ \vec{R} $ is invariant under the GT, so can be set to be zero in any inertial frame,
%  but the relative distance between the boson and the impurity $ \vec{x}- \vec{R}- \vec{v} t $ is not invariant under the GT.
%  It is time-independent in the co-moving frame, but becomes time-dependent in the lab frame.
  Unfortunately, even so, it becomes very difficult to solve Eq.\ref{drivingprime} even in such a static frame.
  It belongs to a quantum impurity problem \cite{kondo} where one need to deal with an impurity scattering on a moving fluid.
  If setting $ g_i =0 $, the system is a gapless system describing by a CFT, then $ g_i \neq 0 $ may act
  as a Boundary condition changing operator in such a boundary CFT. The $ g_i \to \infty $ limit may just directly set
  the Dirichlet  boundary condition $ \psi( \vec{x}= \vec{R} )= 0 $.
  After solving such a quantum impurity problem
  in the static frame, one need to transfer the solution back to the lab frame where it becomes a
  time-dependent again. The earliest theoretical treatment is Landau's original argument by treating
  the impurity as a heavy classical particle and the quantum fluids as classical also which leads to
  $ v_{im[} \sim \Delta/k_0 $.  But such a classical treatment
  may not be precise to describe such a quantum impurity problem ( for a more recent study, see \cite{flowSF}   ).
  It may break down in the presence of lattice  anyway.

{\sl 2. A moving optical lattice }

   Driving the underlying ionic lattice in a solid is hard to achieve in materials, but may be implemented in cold atom systems.
   It is easy to extend a single impurity located at position $ \vec{R} $ to a macroscopic lattice located at
   the ordered array of $ \vec{R}_i $,  so $ V_1( \vec{x} )= \sum_{\vec{R}} v (\vec{x}-\vec{R} ) $ is a single-body
   attractive trapping potential.   We look at the case of driving the lattice at a given velocity $ v $ in the Hamiltonian Eq.\ref{Seq0Nlattice2}:
\begin{eqnarray}
  {\cal H}^{OL}_L & = &  \int d^2 x \psi^{\dagger}( \vec{x} )
   [  -\frac{\hbar^2}{ 2 m} \nabla^2  - \mu- \sum_{\vec{R}} v( \vec{x}- \vec{R} + \vec{v} t ) ] \psi( \vec{x} )
          \nonumber    \\
    & + &  V_{int}
\label{drivingL}
\end{eqnarray}
  where $ \vec{R}_i $ are the initial positions of the ions, $ \vec{v} $ is the driving velocity of the lattice.
  As expected, it is a time-dependent driving lattice system in the lab frame.

  In the frame moving together with the optical lattice, one can set $ \vec{x}^{\prime}= \vec{x}+ \vec{v} t $.
  Then one obtain the Hamiltonian in this co-moving frame ( still drop the $ \prime $ for the notational simplicity ):
\begin{eqnarray}
  {\cal H}^{OL}_{M} & = & \int d^2 x \psi^{\dagger}( \vec{x} )
   [  -\frac{\hbar^2}{ 2 m} \nabla^2  - \mu-  \sum_{\vec{R}} v( \vec{x}- \vec{R} ) -i v \partial_x ] \psi( \vec{x} )
                 \nonumber  \\
    & + &  V_{int}
\label{drivingprimeL}
\end{eqnarray}
  which, as expected, becomes time-independent in the moving frame.

  The most dramatic differences between Eq.\ref{Seq0Nlattice2}  with $ V_1( x )= \sum_{\vec{R}} v (\vec{x}-\vec{R} ) $
  and the current driving lattice case is that
  in the former one boost both the quantum and classical degree of freedoms at the same speed,
  so the relative distance between the boson and any lattice site
  $ \vec{x}^{\prime}- \vec{R}^{\prime}= \vec{x}- \vec{R} $ is invariant under the GT, so it is a time-independent problem in any inertial frame.
  so it can be set to be zero in the tight-binding limit
  $ \vec{x}^{\prime}- \vec{R}^{\prime}= \vec{x}- \vec{R} \to 0 $ in any inertial frame.
  but here $ \vec{R}_i $ are just a array of constants, so invariant under the GT,  but the relative distance between the boson
  and the lattice site $ \vec{x}- \vec{R} + \vec{v} t $ is not invariant under the GT.
  It is time-independent in the co-moving frame, but becomes time-dependent in the lab frame.
  So even one may be able to take the time-binding limit in the former, it breaks down in the latter.
  Due to this fact, it becomes very difficult to solve Eq.\ref{drivingprime} even in such a static frame.
  Then one need to transfer back the solution to the lab frame where it becomes a
  time-dependent problem again. So far, there is not any controlled theoretical treatment on such class of problems.

{\sl 3. A moving straight hard wall along the x-axis  }

   It is easy to extend the macroscopic lattice located at $ \vec{R}_i $ to a continuous wall,
   so $ V_1( \vec{x} )= \int d\vec{R} v (\vec{x}-\vec{R} ) $ with the repulsive interaction $ v (\vec{x}-\vec{R} ) $.
\begin{eqnarray}
  {\cal H}^W_L & = & \int d^2 x \psi^{\dagger}( \vec{x} )
   [  -\frac{\hbar^2}{ 2 m} \nabla^2  - \mu- \int d\vec{R} v( \vec{x}- \vec{R}+ \vec{v} t ) ] \psi( \vec{x} )
              \nonumber   \\
   & + &  V_{int}
\label{drivingLW}
\end{eqnarray}
  where the continuous $ \vec{R} $ are the initial positions of the wall, $ \vec{v} $ is the velocity of the wall.
  As first glance, it seems a time-dependent driving system in the lab frame. A second look finds that by changing the
  variable $ \vec{R}^{\prime} = \vec{R} - \vec{v} t $, it becomes $ \int d\vec{R} v( \vec{x}- \vec{R}+ \vec{v} t )
  =\int d\vec{R}^{\prime}v( \vec{x}-\vec{R}^{\prime} )= V_1( \vec{x} ) $ which is actually time-independent.

  In the frame moving together with the wall, one can set $ \vec{x}^{\prime}= \vec{x}+ \vec{v} t $.
  Then one obtain the Hamiltonian in this co-moving frame ( still drop the $ \prime $ for the notational simplicity ):
\begin{eqnarray}
  {\cal H}^W_{M}  & = & \int d^2 x \psi^{\dagger}( \vec{x} )
   [  -\frac{\hbar^2}{ 2 m} \nabla^2  - \mu- \int d\vec{R}  v( \vec{x}- \vec{R} ) -i v \partial_x ] \psi( \vec{x} )
                             \nonumber  \\
    & + &  V_{int}
\label{drivingprimeLW}
\end{eqnarray}
  which, as expected, becomes time-independent in the co-moving frame.
  Now if one takes the hard wall limit  $ v (\vec{x}-\vec{R} ) \to \infty $ in the co-moving frame,
  then it just sets the boundary condition at the wall $ \psi( \vec{x}= W )=0 $.
  One need to solve Eq.\ref{drivingprimeLW} with such a Dirichlet boundary condition in such a static frame.
  Then one need to transfer back the solution to the lab frame which, as explained above, is still
  a time-independent problem\cite{wire}.
  When the wall is very wide, it is equivalent to the moving SF case discussed in the main text.
  Eq.\ref{drivingLW} and Eq.\ref{drivingprimeLW} formally looks the same, however, the SF phase is a $ U(1) $ symmetry breaking phase where
  even the number of particles is not conserved. Then it gets back to the effective action treatments developed in the main text.
  This is another example of  `` More is different " enriched in \cite{moving}.
  Our new critical velocity Eq.\ref{cvaluenew} presented in the main text
  holds for a extremely wide  channel $ d \to \infty $ where the  Dirichlet boundary condition does not affect the bulk.


\begin{thebibliography}{99}
\bibitem{lattice} It is interesting to compare this relation with that of the finite temperature phase transition
in the SF side of the SF-Mott quantum phase transition in a lattice observed in a moving frame with the velocity $ v $
studied in \cite{moving}: $ \frac{T_c}{T_{c0} }= \sqrt{  1 + (\frac{v}{v_{c}})^2 } > 1 $.
It is the lattice which makes the dramatic difference between the two formulas.
\bibitem{transfer}
 This $ z=2 $ vacancy induced transition can be compared to
 the appearance of the $ z=2 $ transition line in the boosted $ z=1 $ SF-Mott transition presented
 in \cite{moving}. After extracting out the modulation factor, then the $ z=1 $ line transfers to the $ z=2 $ line.
\bibitem{Csymmetry} For example, in the C-symmetry case, the phase and  magnitude in Eq.\ref{comovingSF} are not
conjugate variables anymore. In fact, they become independent, the magnitude mode becomes an independent Higghs mode.
See \cite{response,moving} for details.
\bibitem{groundexcitation} Here, we only use the mean field ground state for the SF or BSF. In fact, as shown in
\cite{BLQHwave} in the context of exciton SF in BLQH and \cite{SOCwave} in the context of a Y-y magnetic state in an fermionic SOC system,
the mean field ground state will be modified by the quantum fluctuations in the ground state which is determined by
the excitation spectrum even at $ T=0 $.
\bibitem{CIC}  It maybe constructive to compare the lattice phonon mode here with that in a lattice mode represented in
\cite{response}, here  $ a $ in $ x \to x+ a $  is any real number,
the $ k_0 $ is also any real number, then $ \theta= k_0 a $ is a continuous number, it comes from a continuous translation symmetry breaking
to a lattice translation normal to the stripe resulting the lattice phnonon mode.
While in the lattice $ x \to x+ a $ where $ a $ must be a lattice constant, $ k_0 a $ is a continuous number only when
$ k_0 $ is an In-commensurate number. it comes from the discrete lattice translation symmetry $ x \to x+ a $ breaking to none
resulting the gapless phason mode.
\bibitem{qz} Here, we use the $ 2+1 $ d notation.
Due to the rotation symmetry around the boost x-axis ( in the $ YZ$ plane ), one may just set $ q^2_y \to q^2_y + q^2_z $ in the 3d He4 case.
\bibitem{RW} In quantum optics, one usually study atom-photon interaction in a frame rotating together with the external pumping frequency,
such a frame is called rotating frame where the atom-photon interaction becomes time-independent.
Unfortunately, so far, there is no work to study how to transfer the time-independent phenomena in the rotating frame back to the lab frame
which is time-dependent.
\bibitem{wire} To make a simple analogy here, a moving charge will produce a time-dependent EM field. but a line of moving charges form
a steady current will produce a static magnetic field, so becomes time-independent.
\end{thebibliography}
\end{document}